# DTLS Performance – How Expensive is Security?

Sebastian Gallenmüller, Dominik Schöffmann, Dominik Scholz, Fabien Geyer, and Georg Carle
*Chair of Network Architectures and Services, Department of Informatics*
*Technical University of Munich*
Garching near Munich, Bavaria, Germany
dtls-performance@list.net.in.tum.de

*Abstract*—Secure communication is an integral feature of many Internet services. The widely deployed TLS protects reliable transport protocols. DTLS extends TLS security services to protocols relying on plain UDP packet transport, such as VoIP or IoT applications. In this paper, we construct a model to determine the performance of generic DTLS-enabled applications. Our model considers basic network characteristics, e.g., number of connections, and the chosen security parameters, e.g., the encryption algorithm in use. Measurements are presented demonstrating the applicability of our model. These experiments are performed using a high-performance DTLS-enabled VPN gateway built on top of the well-established libraries DPDK and OpenSSL. This VPN solution represents the most essential parts of DTLS, creating a DTLS performance baseline. Using this baseline the model can be extended to predict even more complex DTLS protocols besides the measured VPN. Code and measured data used in this paper are publicly available at https://git.io/MoonSec and https://git.io/Sdata.

Transport layer security (TLS) is the workhorse of today's encryption protocols, securing services such as e-mail or web communication. In 2002, TLS was standardized for protocols such as TCP. As TLS itself relies on the reliability of the underlying protocol, TLS cannot be trivially applied to protocols lacking this capability such as UDP. Datagram transport layer security (DTLS) – standardized in 2006 [1], revised in 2012 [2] – solves this shortcoming, enabling secure communication over UDP-based protocols. There are various applications utilizing DTLS, for instance, WebRTC which is implemented in modern browsers to allow latency critical, DTLS-secured VoIP connections over the Secure Real-Time Transport Protocol (SRTP). Another area of application is the domain of the Internet of Things (IoT) and specialized protocols such as the Constrained Application Protocol (CoAP). Furthermore, DTLS can be used for tunneling protocols, offering a simple and encrypted service, but also lower reliability.

Every DTLS-enabled application involves a number of essential processing steps: packet IO, tracking of state for different connections, and packet processing which includes the encryption/decryption of packets. In this paper, we investigate a prototype VPN gateway application which involves these *essential processing steps*. At the same time, this basic VPN application allows measuring the effects of DTLS in its pure form, unaffected by higher layer protocols. We call our application MoonSec. MoonSec's modular architecture allows benchmarking the processing steps separately to quantify the individual costs of each step. As these processing steps must be performed by every DTLS-enabled application, the benchmarks presented in this paper are part of higher level protocols utilizing DTLS. By adding application-specific processing steps our model can be extended to predict the performance of these higher level protocols.

Our paper offers the following key contributions:
- a generic model to describe the performance of DTLS applications,
- an extension of this model to reflect the characteristics of a DTLS VPN gateway,
- the modular DTLS-based VPN gateway MoonSec, sharing its fundamental building blocks with more complex DTLS applications, and
- measurements to quantify the effect of basic network characteristics and the chosen security parameters alike.

The paper is structured as follows: Section I investigates related work. A performance model for describing packet processing applications is introduced in Section II, which is refined to reflect the specific needs of a VPN gateway. Section III introduces the DTLS VPN gateway MoonSec and its measurements in Section IV where also the model is validated. A short remark about the reproducibility of our experiments is presented in Section V, before concluding the paper in Section VI.

## I. RELATED WORK

RFCs 5246 [3] and 6347 [2] present the DTLS protocol, which is used across different domains:

One such protocol is WebRTC, which is included in every current major browser implementation, and uses DTLS to establish keys. SRTP uses these keys for the data transmission [4]–[6]. SRTP offers a service for applications like video conferencing where simplicity and security is more important than a continuous error-free data transmission. In light of this, DTLS provides a service better suited for this purpose compared to traditional TLS.

IoT is an emerging market with a need for secure protocols. Such devices often rely on UDP for communication, making DTLS the natural choice for encryption, thus eliminating the need for a TCP/TLS stack [7]. Resource consumption can be further optimized by using specialized cipher suites and pre-shared keys [8], making DTLS an attractive protocol for such devices.

VPNs are another application based on DTLS which usually model the "roadwarrior" scenario, but might also be used for site-to-site traffic. Tunneling over a reliable connection can result in a TCP-over-TCP situation, which is shown to deteriorate the performance of the tunneled protocol [9]. There already exist multiple VPN solutions utilizing UDP, however, none were usable for the measurements in the scope of this work. OpenVPN is a VPN, which can be used over UDP, but internally uses TLS and ESP [10], [11]. DTLS is originally designed to run on its own without requiring mechanisms like TLS or ESP. In this paper, we want to investigate DTLS on its own, therefore we did not further consider the OpenVPN solution. Another widely used VPN solution is Cisco's AnyConnect, for which the open source implementation OpenConnect exists. However, OpenConnect utilizes a pre-release version of DTLS [12], [13]. We decided that it is best practice to stick with established and final standards. Therefore, we developed a custom DTLS application, MoonSec, which relies on the standardized version of DTLS. MoonSec tries to resemble a generic VPN secure gateway relying on well-established libraries such as OpenSSL implementing DTLS as described by RFC 6347 [2] (see Section III). Furthermore, DTLS is designed to run in userspace [2], which makes DTLS an excellent fit for high-performance userspace packet processing frameworks such as DPDK used by MoonSec (see Section III).

Another UDP-based VPN is WireGuard [14]. MoonSec and WireGuard have similar design goals; both are optimized for performance and can run in userspace. WireGuard only supports the ChaCha20 cipher with Poly1305 [15]. This cipher suite performs well in software-only implementations, almost tripling performance compared to AES in non-accelerated environments [16]. Due to its beneficial properties, this cipher suite was also integrated into DTLS [17]. Limiting the cipher suites and possible configurations for WireGuard is done intentionally lowering complexity, simplifying configuration, and securing the VPN at the same time. In contrast to WireGuard, DTLS offers the possibility for different cipher suites and configurations to be tailored to the specific needs of an application. This high degree of freedom for configuration of DTLS requires a careful evaluation of security properties. RFC 7525 [18] gives recommendations for the secure usage of (D)TLS.

Various studies investigate the performance of DTLS-based protocols and their security overhead. A survey by Subramanian and Dutta [19] demonstrates the costs of SRTP compared to unsecured RTP, with higher costs in terms of CPU usage, memory usage, and connection setup time. Alexander et al. [20] perform an in-depth analysis of SRTP for VoIP, comparing different cipher suites and their influence on latency and jitter. Both presented studies try to measure overhead for secure communication in a specific domain. We provide basic measurements which will be used to derive a generic model that can be adapted to a broad range of domains and applications.

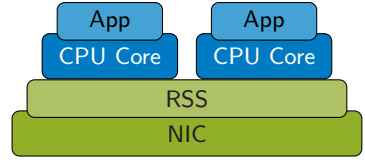

Figure 1: Inter-core scalability architecture

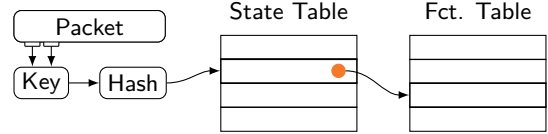

Figure 2: Application design

## II. MODELING

We propose a generic model for describing stateful packet processing applications, which can handle different packet flows individually. To construct a model, we first identify the architectures and design principles of such packet processing applications. This allows us to characterize the capacity of these applications for handling different flows, which we call scalability. We define two levels of scalability: *inter-core* across different CPU cores and *intra-core* on a single CPU core.

### A. Inter-core Scalability

Typical high-performance packet processing applications are scalable across many cores to leverage modern multi-core architectures depicted in Figure 1. Received packets are distributed in hardware to different cores via a feature called receive side scaling (RSS). RSS can employ filters, such as 5-tuple hashing, to maintain a fixed flow-to-core mapping. If the threads of such an application avoid information sharing, which subsequently would require synchronization mechanisms, these applications can be scaled with hardly any performance penalties. The same technology was used by Emmerich et al. in the MoonRoute [21] project, which implements a high-performance software router on top of DPDK. Similarly, mTCP [22], a high-performance user-space TCP stack tries to keep as much data as possible local to one core, such as flow hash tables, in order to counter inter-core contention. Given the right architecture, MoonRoute and mTCP show that almost perfect linear scaling with the number of CPU cores can be achieved: twice as many flows can be processed if the number of CPU cores is doubled.

### B. Intra-core Scalability

Stateful flow handling requires state keeping for individual flows. A basic architecture for handling flows on a single core is shown in Figure 2. This architecture involves a data structure for per-flow state tracking, which we call state table. The state can be accessed by extracting a key from a received packet such as a 5-tuple for TCP/UDP packets. Protocols, such as QUIC, may provide a special identifier header field which can directly be used for flow identification. A second table, called function table, contains all functions for packet processing. The function table is indexed using the current state of the packet/flow as determined by the lookup in the state table. Packet processing works as follows: the packet is received, the current state is extracted from the state table, then the function is executed with the packet and the current state as arguments.

*C. VPN Model*

To describe the overall performance of an application, inter-core scalability and intra-core scalability must be taken into account. If an application follows the architecture presented in Section II-A, the performance scales linearly with the number of cores, which makes modeling the inter-core scalability trivial. To determine the overall performance of our application, we additionally have to determine the intra-core scalability, i.e., the performance of a single core. Following the application presented in Figure 2, we define four basic processing steps: packet reception & transfer, key extraction (e.g., 5-tuple) & hashing, state table handling, and packet processing. For each of these steps, the costs in terms of CPU cycles are determined.

*a) Packet Reception & Transfer:* Gallenmüller et al. [23] describe a model for packet IO on userspace packet processing frameworks such as DPDK or netmap. Our model follows the same approach for reception and transfer of packets with the cost depending on the number of packets processed instead of packet size. This scales linearly with the number of packets processed.

*b) Hashing:* Hashing the 5-tuple of flows is modeled as a separate step and not included in the state access step because of its importance on the VPN application we want to model. Choosing a non-cryptographic hash function would allow for an attacker to induce collisions in the state table. This kind of attack is also referred to as "hash flooding" [24]. Collision handling would cause additional overhead for the next processing step – the handling of the state table – which offers a potential basis for an attack on the VPN application. The strength of the hash function is, therefore, an essential feature of the overall VPN security, which justifies modeling the hash function as a separate element.

*c) State Table Handling:* Computational complexity of this step includes memory access. This can be modeled as two separate steps: memory *allocation* and memory *access*. *Allocation*, in this context, involves reserving the memory and inserting an element into the state table. State allocation is done once when a new connection is created. After that, no additional state is allocated for this connection. The memory is only accessed and updated, but not extended. Therefore, the overall cost of state table handling heavily depends on the usage scenario: short-lived connections are dominated by allocation costs happening at the creation of a connection, whereas access costs prevail in long-living connections with a large number of packets.

*d) Packet Processing:* The main task of this step is the decryption or the encryption of the packet data. Encryption algorithms have an inherent cost depending on the kind of algorithm used. We model two different classes of encryption and decryption costs – costs during normal operation and costs for an initial connection setup. VPNs often require an initial key exchange for a connection setup. IPsec with IKEv2 [25], and DTLS [2] are two examples utilizing an initial public key handshake. However, both protocols also support modes employing pre-shared keys avoiding public key cryptography and therefore the costs for the connection setup entirely.

## III. ARCHITECTURE OF MOONSEC

The architecture of MoonSec is inspired by the model created from the four building blocks presented in the previous section. We selected well-established frameworks and technologies: to handle the packet IO we use DPDK [26]; for 5-tuple hashing packets we use SipHash-2-4 [24]; we integrated Google's DenseMap [27] as our state table; and we rely on OpenSSL for encryption and decryption.

DPDK implements kernel-bypass technology accelerating packet processing in userspace. From several frameworks available, MoonSec relies on DPDK because of its maturity and the high performance [23]. Following DPDK's design paradigm MoonSec entirely runs in userspace.

As explained in Section II-C, a cryptographic hash function is required for effective mitigation of hash flooding attacks. The second requirement is high performance as the hashing is performed once per incoming packet (e.g., up to 14.88 Mpps for 10 Gigabit Ethernet line-rate). A high-performance hash function specifically designed for short inputs like packets is SipHash-2-4 [24]. SipHash is integrated with various software and programming languages (e.g., Python, Rust, Ruby). Other hashing algorithms, such as SpookyHash or City hash, may achieve higher performance but offer lower collision resistance which disqualifies them from being used in MoonSec.

Since connection handling involves a hash table lookup, selecting an efficient implementation is crucial for MoonSec's overall performance. Google's DenseMap has the best performance among the tested algorithms. These tests were confirmed by benchmarks done by Tessil [28].

Besides its own VPN solution, relying on TLS and ESP (see Section I), OpenSSL offers an RFC-compliant DTLS implementation. We chose OpenSSL [29] because it is a mature, widely deployed cryptographic library, performing the cryptographic operations of MoonSec.

The major improvements of MoonSec compared to the state of the art are: *RFC-compliance*, a focus on *high performance*, and a *flexible, modular architecture* constructed of four basic building blocks.

## IV. PERFORMANCE EVALUATION

This section presents the test setup, test design and results of our performance evaluation of MoonSec. Results are used to create a performance model describing the costs of DTLS-based protocols.

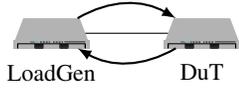

Figure 3: Setup of the test

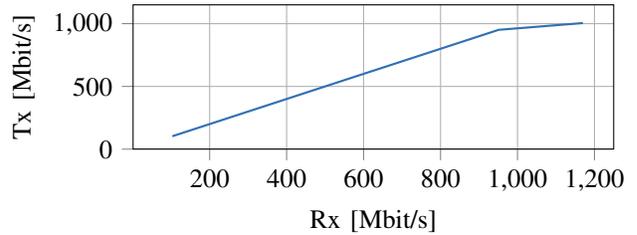

Figure 4: Resulting throughput (Tx) of the MoonSec DuT for a given offered load (Rx)

## A. Setup and Methodology

The setup consists of two bidirectionally connected servers as shown in Figure 3. One is the Device-under-Test (DuT), running MoonSec, and the second server is used as load generator and sink.

Both servers feature Intel XL710 dual port NICs supporting 40 Gbit/s. The DuT has one Intel E5-2620 v3 processor, which has 6 physical and 12 logical cores with a base clock frequency of 2.40 GHz, a turbo of 3.2 GHz, and 15 MB of L3 cache. AES hardware acceleration (AES-NI) is disabled on the DuT for all experiments.

All tests are conducted using Debian images with Linux kernel version 4.14.0. Variables were the number of connections, the amount of traffic per connection, and the used cipher suite. Every connection follows the standard DTLS – establishing keys using a handshake before sending traffic.

The MoonSec framework is configured to pass encrypted payloads to OpenSSL, which decrypts the content, after which it is then re-encrypted and sent back to the load generator. Using this approach we model a DTLS VPN security gateway (SG), which passes equal amounts of traffic in both directions. The CPU time of functions is measured using time stamp counter (`TSC`) of the CPU [30]. MoonSec follows the design presented in Section II-A. Therefore, we assume linear multi-core scaling, as each thread is isolated, sharing no information between cores. To determine the intra-core scalability (see Section II-B), we limit MoonSec to only use a single core for this evaluation. All experiments were run on a single core of the DuT clocked with the turbo frequency of 3.20 GHz. The payload data was 500 B per packet. This size was chosen to be below 576 B, the minimum packet size which must be accepted by any receiver accepting IPv4 [31]. This avoids fragmenting IP packets which could impact measurement results because of fragmentation handling.

## B. Measurements

Two types of tests were conducted. First, we treat the DuT as a black box, measuring only the resulting throughput. Second, the individual cost of different modules within the VPN is measured and modeled.

*1) Black Box Measurements:* The result of our throughput measurement is shown in Figure 4, depicting the amount of traffic the DuT is able to process. All measurements are performed using a single connection. The cipher suite was AES256-GCM using an ECDHE handshake. Throughput scales linearly with the offered traffic up to 950 Mbit/s. Afterward, the capacity of MoonSec is exceeded, resulting in packet drops.

*2) White Box Measurements:* The white box measurements use the following input parameters:

- c: Number of connections
- b: Number of bytes transmitted
- p: Number of packets transmitted

There are information criteria available to compare different models, such as the AIC [32]. Models with a lower number of parameters are preferred over more complex models. However, the models must not lose their predictive power due to over-simplification. Following this idea, we show that our model can predict the performance using only the three given input parameters. To assess the prediction quality of our model, we measure the difference between real and predicted values expressed by the symmetric mean absolute percentage error (sMAPE) [32].

Our model measures costs in CPU cycles. Applying these costs onto a given CPU cycle budget allows a throughput prediction for such a system. Other system bottlenecks such as the available Ethernet bandwidth may further limit the throughput. Such limits can be represented as fixed bounds for performance in addition to our model prediction. For all measurements presented in this paper the CPU resources were the only limiting factor of packet processing.

*a) Packet Reception & Packet Transfer:* As explained in Section III, MoonSec uses the userspace packet processing framework DPDK for network access. Utilizing such a framework ensures constant costs for packet IO, i.e., costs depend on the number of packets processed, not on their size. Literature suggests a performance of roughly 100 cycles per IO operation, including reception and transfer [23], [33]. We measured reception and transfer separately, showing slightly higher costs. The average reception of a packet costs 77 CPU cycles, transfer costs 66 CPU cycles. We attribute the increased costs to differences in hardware and different DPDK versions.

$$\text{tx}(p) = p \cdot 66 \quad (1)$$
$$\text{rx}(p) = p \cdot 77 \quad (2)$$

Equations 1 and 2 show the average cost of packet transmission and reception. As per-packet costs are constant, both equations have p as the sole input parameter. The performance figures for userspace packet processing frameworks, such as DPDK, tend to be stable, rendering the calculation of an sMAPE value redundant [33].

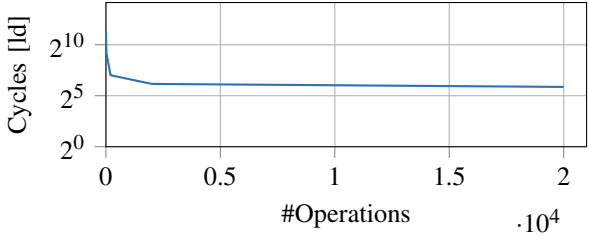

Figure 5: Average costs per hashing operation (SipHash)

*b) Hashing:* Figure 5 depicts the costs of the SipHash-2-4 algorithm. The hashing itself is done over a 5-tuple (source/destination IP, source/destination port, protocol type), which is used to map packets to connections. After an initial phase with higher costs, SipHash converges to 62 CPU cycles per hashing operation. As hashing is done once per incoming packet the overall hashing costs can be calculated by equation 3, with p representing the number of incoming packets. 5-tuple hashing is independent of packet size, leading to constant per-packet costs.

$$\text{hash}(p) = p \cdot 62 \qquad (3)$$

*c) State table Handling:* Google's DenseMap is used for the state tracking. Since it is an integral part of MoonSec, we developed a model for its performance. The two important operations in this context are insertions and lookups. Per the hash map's implementation, the map grows regularly if a certain fill state is reached, resulting in a sawtooth pattern using a log scale for state insertions, depicted in Figure 6. A trend towards a converging behavior can be observed.

$$\text{mem}(c) = c \cdot 354 + 1477 \qquad (4)$$

$$\text{dMapS}(c) = c \cdot \left(400 + 170 \cdot \left(\frac{2^{\lfloor \log_2(c) \rfloor + 1} - 8}{c} - 1\right)\right) \qquad (5)$$

(sMAPE:15%)

Equation 5 fits the sawtooth pattern of the insertion cost. It consists of static costs (400 cycles) plus the dynamic costs of the sawtooth pattern on top of it. The $\log_2$ portion of the equation stems from the growth behavior, i.e., the map doubles its size every time it is at least half filled. Equation 5 is not correct for a low number of connections and should therefore only be applied to 1000 connections or more. The number of connections c is the sole input parameter of this equation, representing the setup cost happening once per connection.

In addition to the setup costs in the hash table, we measured the memory allocation for the state itself. The average costs for this operation are modeled in Equation 4. There are applications which preallocate the entire memory before starting normal operation. If such an application design is chosen, the memory allocation costs can be ignored during operation.

$$\text{denseMapR}(p) = p \cdot 118 \qquad (\text{sMAPE:18\%}) \qquad (6)$$

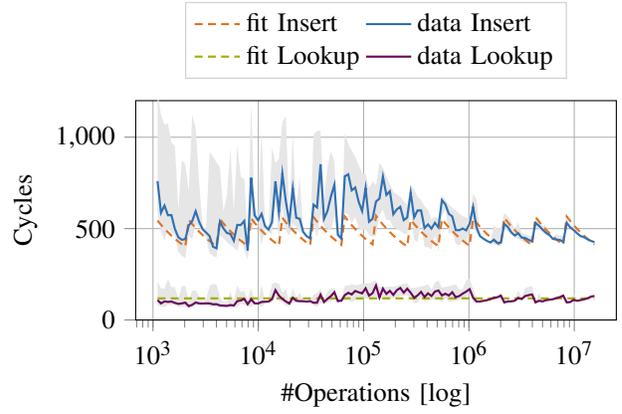

Figure 6: Average costs per operation performed in Google DenseMap. Minimum and maximum values are visualized using the gray area.

Similarly, the lookup, as modeled in Equation 6, has a constant cost, as is expected for hash tables. On average 118 CPU cycles were needed to access an entry in the state table.

*d) Packet Processing:* The first step of establishing a DTLS connection is a handshake operation, which uses either modp-based Diffie-Hellman Ephemeral keys (DHE) or elliptic-curve based Diffie-Hellman Ephemeral keys (ECDHE). We used three different encryption algorithms: AES 128, AES 256, and ChaCha20[1].

A comparison between these algorithms is depicted in Figure 7a, with per-packet costs divided into the four building blocks of our model. For this measurement, we used a single connection transferring a total of 1 kB of payload. With only very little data transmitted the costs shown in Figure 7a are consisting almost entirely of the handshake operation. These costs are part of the OpenSSL block, with the other costs being almost invisible. The figure also shows that modp DHE is about four times more expensive than ECDHE. Figure 7b shows the behavior of OpenSSL if 10 MB data are transferred over the connection. By transferring a large amount of traffic per connection, the cost of the handshake is amortized over the duration of the entire connection. Although a difference between the modp and the ECDH key exchange is still visible, it is not significant (Note that the scaling of the y-axis between Figures 7a and 7b differs by a factor of 100). The fastest cipher is ChaCha20, followed by AES 128 bit and AES 256 bit. It is surprising, that AES 256 is only marginally more expensive compared to AES 128, whilst working on the doubled key size. ChaCha20 was designed as a faster alternative to AES in pure software implementations [15]. It should be noted, that AES-NI was disabled, therefore AES, as well as ChaCha20 ran entirely in software, without special hardware support.

While the handshake is very costly, it does not have fixed

---

[1] The used cipher suites are DHE-RSA-AES128-GCM-SHA256, ECDHE-RSA-AES128-GCM-SHA256, DHE-RSA-AES256-GCM-SHA384, ECDHE-RSA-AES256-GCM-SHA384, DHE-RSA-CHACHA20-POLY1305, and ECDHE-RSA-CHACHA20-POLY1305.

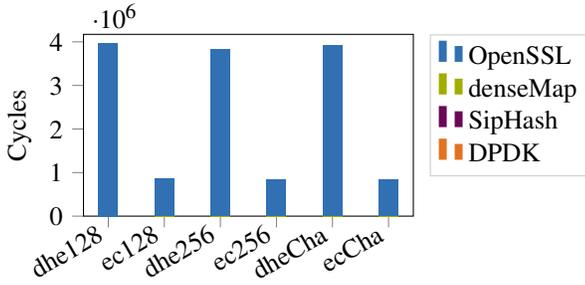

(a) 1 kB transmitted

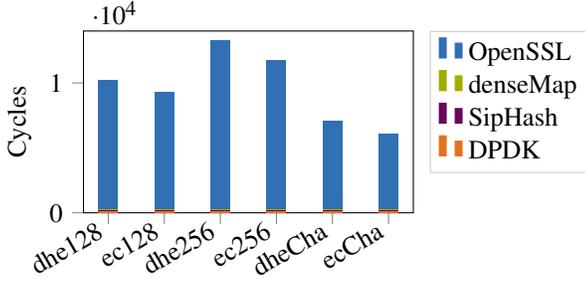

(b) 10 MB transmitted

Figure 7: Average per-packet costs depending on crypto algorithm

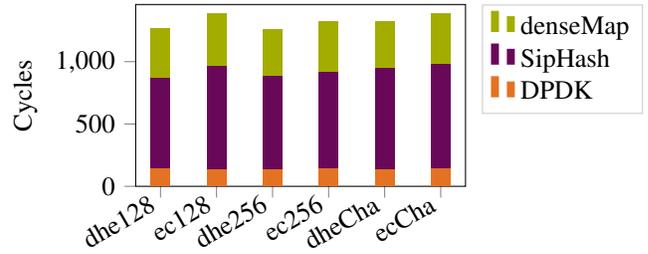

(a) 1 kB transmitted

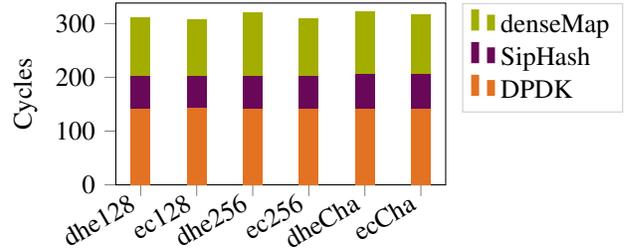

(b) 10 MB transmitted

Figure 9: Average per-packet costs depending on crypto algorithm without OpenSSL

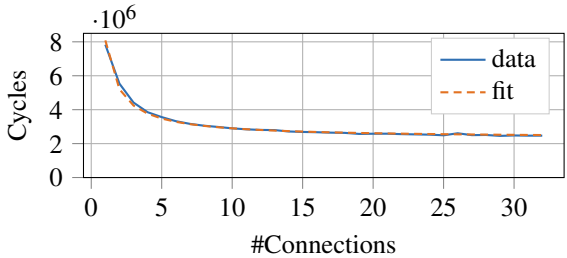

Figure 8: Average costs per connection while performing ECDHE using OpenSSL

per-connection costs, at least not for ECDHE. Figure 8 shows the cost of the ECDHE handshake in relationship to the number of connections. It can be seen, that the costs are declining with a growing number of connections. The rationale is, that at first, a key needs to be generated, which is a costly operation, which will later be reused for multiple connections. Reusing the same key pair in ECDHE operations is part of OpenSSL's DTLS implementation. The differences between the individual cipher suites are overshadowed by the high costs of ECDH and therefore neglected in Figure 8.

The above findings are modeled in Equation 7, for the handshake, and Equation 8 for the encryption during the established connection. The CPU costs for Equation 7 are taken from Figure 8 which results in the one-time costs for ECDH key pair generation of 5759960 CPU cycles. A per-connection overhead of 2325634 CPU cycles is measured. The overall costs for the connection setup depend on the number of connections c. Equation 8 uses ECDHE and ChaCha20 as cipher suites since they were found to be the fastest. The average costs for these cipher suites are 12 CPU cycles per byte. Costs for encryption are constant per byte and depend on the amount of data to encrypt, i.e., b. These equations can trivially be adapted to other ciphers applying the numbers of Figure 9.

$$\text{openSSlS}(c) = 5759960 + c \cdot 2325634 \quad \text{(sMAPE: 1.5\%)} \tag{7}$$

$$\text{openSSlR}(b) = 12 \cdot b \tag{8}$$

*e) Model validation:* Figure 9 shows the transmission without the costs of OpenSSL, to make the other costs visible. It is worth noting, that the cipher has no effect on the underlying framework beyond a typical margin of measurement error. The total non-cryptographic overhead is about 300 cycles per packet for long running connections (see Figure 9b) and non-cryptographic, per-packet costs of roughly 1100 cycles (see Figure 9a) for short-lived connections. All previous measurements (excluding Figures 7, 8, and 9) investigated IO, hashing and the state table handling separated from the other parts of the VPN application. Here we measured the application with all parts combined, which we use in the following to validate the proposed model:

DPDK has the same costs as predicted by Equations 1 and 2 of roughly 130 cycles. The costs for SipHash differ between Figures 9a and 9b. The reason for this difference is the number of iterations executed in the different scenarios. With roughly 500 cycles and 100 cycles, the results are in line with Figure 5. The costs for DenseMap in Figure 9b show a running connection. Such costs are correctly described by Equation 6. Figure 9a shows mainly the costs for a single state allocation. As previously explained, Equation 5 does not work for small map sizes. We assume that for small map sizes CPU caches have a large impact on performance, therefore we could not identify a regular pattern.

*f) Comprehensive model:* The model can be simplified by combining the per-packet costs expressed by Equations 1, 2, 3, and 6, resulting in overall per-packet costs of 323 cycles per packet. Given an average packet size the encryption costs can also be converted to only depend on p. For our setup we used 576-byte packets with a 500-byte payload. Assuming costs of 12 cycles per byte payload (see Equation 8) leads to encryption costs of 6000 cycles per byte. The overall per-packet costs are described by Equation 9:

$$\text{overallPerPacket}(p) = 6323 \cdot p \qquad (9)$$

For the per-connection costs we combine Equations 4, 5, and 6. To simplify Equation 5, we always assume the worst case of the sawtooth pattern leading to costs of 570 cycles per connection. The overall per-connection costs are expressed in Equation 10:

$$\text{overallPerCon}(p) = 5761437 + 2326558 \cdot c \qquad (10)$$

The final model is expressed by Equation 11, which only requires two input parameters to predict the performance of MoonSec.

$$\text{overallC}(c, p) = 5761437 + 2326558 \cdot c + 6323 \cdot p \qquad (11)$$

Note that Equation 11 is only valid for long runtimes involving more than 1000 connections. For fewer or short-lived connections the parameters have to be adapted. This equation emphasizes the high initial costs for connection setup compared to the lower per-packet costs. For efficient resource usage connections should always be long running to amortize these costs.

## C. Value & Limits of the model

Equation 11 depicts a clear trend demonstrating that the per-connection overhead surpasses the per-byte costs by a factor of 368. These parameters were determined through a series of experiments performed on an Intel CPU which limits the predictive power of our model to this specific architecture. However, we consider the factor to be big enough to demonstrate a general trend – the per-connection setup costs are considerably more expensive than the per-packet costs. Therefore, we assume this trend to be true even for other CPU architectures as it is rather a result of the performed source code than the CPU architecture in use. To transfer our model to different target architectures, our benchmarks have to be repeated on these platforms. After that, the CPU-specific costs for the target platform are known and the model can be adapted. To simplify this process, we release MoonSec as open source including measurements and plotting scripts.

## V. REPRODUCIBILITY

As part of an ongoing effort to foster reproducible research in computer science we want to release our data to interested parties: the source code is licensed under open sources licenses (BSD-3, MIT), and available on https://git.io/MoonSec. In addition to the source code, the measured data which was analyzed in this paper is made available on https://git.io/Sdata including the scripts for generating all plots used in this paper.

## VI. CONCLUSION

In this paper, we presented a DTLS performance model consisting of four basic building blocks describing the main cost factors of a DTLS-based application – packet IO, cryptographic hashing, state handling, and cryptographic processing costs. We implemented a high-performance VPN, MoonSec, to validate the modeled findings. Applying our model, we identified the cryptographic operations being the main cost factor. There is a measurable influence of the cryptographic algorithm on the resulting costs. Costs can be halved between the most expensive and the least expensive of the investigated state-of-the-art encryption algorithms at least for non-hardware accelerated ones. Handshakes are even costlier, with the modp-based DHE ciphers almost quadrupling the costs of the more cost-efficient elliptic curve based DHE ciphers. Therefore, we recommend using elliptic curve based cipher suites with ChaCha20 as an encryption algorithm. The performance improvement can be significant, especially considering the high relative costs of the handshake for short-lived connections.

Beneath the encryption software layer, the powerful VPN application MoonSec was built, relying on state-of-the-art frameworks and libraries. This ensures the model to reflect the high performance which can be achieved by modern applications. This framework features a scalable, multi-threaded architecture running on top of DPDK, using DenseMap for efficient connection state tracking, while mitigating hash flooding attacks by implementing the cryptographic SipHash. OpenSSL, a well-known cryptographic library supporting DTLS, handles encryption and decryption.

Our measurements were performed on powerful servers in high-load scenarios. However, DTLS is applied across different domains. In other domains, additional resource constraints beyond mere CPU usage may apply. For future work, we want to extend our model to reflect different CPU architectures involving additional parameters, e.g., memory consumption or energy consumption, which are highly relevant for DTLS target domains like mobile devices or IoT applications.

## ACKNOWLEDGMENT

This work was supported by the DFG Priority Programme 1914 Cyber-Physical Networking, the High-Performance Center for Secure Networked Systems, and by the German-French Academy for the Industry of the Future.

## REFERENCES

[1] E. Rescorla and N. Modadugu, "Datagram Transport Layer Security," RFC 4347, Apr. 2006. [Online]. Available: https://rfc-editor.org/rfc/rfc4347.txt

[2] ——, "Datagram Transport Layer Security Version 1.2," 2012. [Online]. Available: https://tools.ietf.org/html/rfc6347
[3] T. Dierks and E. Rescorla, "The Transport Layer Security (TLS) Protocol Version 1.2," 2008. [Online]. Available: https://tools.ietf.org/html/rfc5246
[4] M. Baugher, D. A. McGrew, M. Naslund, E. Carrara, and K. Norrman, "The Secure Real-time Transport Protocol (SRTP)," RFC 3711, 2014. [Online]. Available: https://tools.ietf.org/html/rfc3711
[5] D. A. McGrew and E. Rescorla, "Datagram Transport Layer Security (DTLS) Extension to Establish Keys for the Secure Real-time Transport Protocol (SRTP)," 2010. [Online]. Available: https://tools.ietf.org/html/rfc5764
[6] E. Rescorla, "WebRTC Security Architecture (Draft 14)," 2018. [Online]. Available: https://tools.ietf.org/html/draft-ietf-rtcweb-security-arch-14
[7] S. Raza, L. Seitz, D. Sitenkov, and G. Selander, "S3k: Scalable security with symmetric keys—dtls key establishment for the internet of things," *IEEE Transactions on Automation Science and Engineering*, vol. 13, no. 3, pp. 1270–1280, 2016.
[8] H. Tschofenig and P. Eronen, "Pre-Shared Key Ciphersuites for Transport Layer Security (TLS)," RFC 4279, Dec. 2005. [Online]. Available: https://rfc-editor.org/rfc/rfc4279.txt
[9] O. Honda, H. Ohsaki, M. Imase, M. Ishizuka, and J. Murayama, "Understanding TCP over TCP: effects of TCP tunneling on end-to-end throughput and latency," in *Performance, Quality of Service, and Control of Next-Generation Communication and Sensor Networks III*, vol. 6011. International Society for Optics and Photonics, 2005, p. 60110H.
[10] OpenVPN Developers, "Openvpn security overview," 2 2018. [Online]. Available: https://openvpn.net/index.php/open-source/documentation/security-overview.html
[11] ——, "Why openvpn," 4 2018. [Online]. Available: https://openvpn.net/index.php/open-source/335-why-openvpn.html
[12] OpenConnect Developers, "Openconnect," 2 2018. [Online]. Available: http://www.infradead.org/openconnect/index.html
[13] ——, "Openconnect anyconnect implementation details," 2 2018. [Online]. Available: http://www.infradead.org/openconnect/anyconnect.html
[14] J. A. Donenfeld, "Wireguard: Next Generation Kernel Network Tunnel," in *Proceedings of the 2017 Network and Distributed System Security Symposium, NDSS*, vol. 17, 2017.
[15] D. J. Bernstein, "Chacha, a variant of salsa20," in *Workshop Record of SASC*, vol. 8, 2008, pp. 3–5.
[16] Y. Nir and A. Langley, "ChaCha20 and Poly1305 for IETF Protocols," RFC 7539, May 2015. [Online]. Available: https://rfc-editor.org/rfc/rfc7539.txt
[17] A. Langley, W.-T. Chang, N. Mavrogiannopoulos, J. Strombergson, and S. Josefsson, "ChaCha20-Poly1305 Cipher Suites for Transport Layer Security (TLS)," RFC 7905, Jun. 2016. [Online]. Available: https://rfc-editor.org/rfc/rfc7905.txt
[18] Y. Sheffer, R. Holz, and P. Saint-Andre, "Recommendations for Secure Use of Transport Layer Security (TLS) and Datagram Transport Layer Security (DTLS)," 2015. [Online]. Available: https://tools.ietf.org/html/rfc7525
[19] S. V. Subramanian and R. Dutta, "Comparative study of secure vs. non-secure transport protocols on the sip proxy server performance: An experimental approach," in *Advances in Recent Technologies in Communication and Computing (ARTCom), 2010 International Conference on*. IEEE, 2010, pp. 301–305.
[20] A. L. Alexander, A. L. Wijesinha, and R. Karne, "An Evaluation of Secure Real-time Transport Protocol (SRTP) Performance for VoIP," in *Network and System Security, 2009. NSS'09. Third International Conference on*. IEEE, 2009, pp. 95–101.
[21] P. Emmerich, S. Gallenmüller, R. Schönberger, D. Raumer, and G. Carle, "Architectures for fast and flexible software routers."
[22] E. Jeong, S. Wood, M. Jamshed, H. Jeong, S. Ihm, D. Han, and K. Park, "mtcp: a highly scalable user-level TCP stack for multicore systems," in *11th USENIX Symposium on Networked Systems Design and Implementation (NSDI 14)*. Seattle, WA: USENIX Association, 2014, pp. 489–502. [Online]. Available: https://www.usenix.org/conference/nsdi14/technical-sessions/presentation/jeong
[23] S. Gallenmüller, P. Emmerich, F. Wohlfart, D. Raumer, and G. Carle, "Comparison of Frameworks for High-Performance Packet IO," in *ACM/IEEE Symposium on Architectures for Networking and Communications Systems*, 2015.
[24] J.-P. Aumasson and D. J. Bernstein, "SipHash: a fast short-input PRF," in *International Conference on Cryptology in India*. Springer, 2012, pp. 489–508.
[25] C. Kaufman, P. Hoffman, Y. Nir, P. Eronen, and T. Kivinen, "Internet Key Exchange Protocol Version 2 (IKEv2)," 2014. [Online]. Available: https://tools.ietf.org/html/rfc7296
[26] DPDK Authors, "Data Plane Developement Kit," 2018. [Online]. Available: http://dpdk.org/
[27] Google, "Sparsehash github repository," 2016. [Online]. Available: https://github.com/sparsehash/sparsehash
[28] Tessil, "Benchmark of major hash maps implementations," 2017. [Online]. Available: https://tessil.github.io/2016/08/29/benchmark-hopscotch-map.html
[29] O. Developers, "OpenSSL Website," 2018. [Online]. Available: https://www.openssl.org/
[30] G. Paoloni, "How to benchmark code execution times on intel ia-32 and ia-64 instruction set architectures," *Intel Corporation*, p. 123, 2010.
[31] J. Postel, "Internet Protocol," RFC 791, Sep. 1981. [Online]. Available: https://rfc-editor.org/rfc/rfc791.txt
[32] T. Chis, "Performance Modelling with Adaptive Hidden Markov Models and Discriminatory Processor Sharing Queues," July 2016.
[33] L. Rizzo, "netmap: a novel framework for fast packet I/O," in *USENIX Annual Technical Conference*, 2012.